\documentclass[11pt]{article}
\usepackage{graphicx}
\usepackage{amssymb}
\usepackage{epstopdf}
\newcommand{\jindex}{j_{(i)}}
\newcommand{\kdex}{k}
\font\num=msbm10

\title{The Generalised Liar Paradox: A Quantum Model and Interpretation}
\author{Diederik Aerts, Jan Broekaert and Bart D'Hooghe\\ \\
        \normalsize\itshape
        Center Leo Apostel for Interdisciplinary Studies (CLEA) \\
         \normalsize\itshape
         Foundations of the Exact Sciences (FUND), Department of Mathematics \\
        \normalsize\itshape
        Vrije Universiteit Brussel, 1160 Brussels, 
       Belgium \\
        \normalsize
        E-Mail: \textsf{diraerts@vub.ac.be, jbroekae@vub.ac.be} \\
\textsf{bdhooghe@vub.ac.be}
		}
\date{}
\begin{document}
\maketitle
\begin{abstract}
\noindent
The formalism of abstracted quantum mechanics is applied in a model 
of the generalized Liar Paradox. Here, the Liar Paradox,  a 
consistently testable configuration of logical truth properties,  is 
considered a dynamic conceptual entity in the cognitive sphere 
(Aerts, Broekaert, \& Smets, 1999, 2000, Aerts, Broekaert, \& Gabora 1999, 
2000, 2002).  Basically, the intrinsic contextuality of the 
truth-value of the Liar Paradox is appropriately covered by the 
abstracted quantum mechanical approach.   The formal details of the 
model are explicited here for the generalized case. 
We prove the 
possibility of constructing  a quantum model of the $m$-sentence 
generalizations of the Liar Paradox. This includes (i) the 
truth-falsehood state of the m-Liar Paradox can be represented by an 
embedded $2m$-dimensional quantum vector in a $(2m)^m$ dimensional 
complex Hilbert space, with cognitive  interactions  corresponding to 
projections, 
(ii) the construction of a continuous `time' dynamics 
is possible: typical  truth and falsehood value oscillations are 
described by Schr\"odinger evolution,
(iii) Kirchoff  \& von Neumann 
axioms are satisfied by introduction of `truth-value by inference' 
projectors, (iv) time invariance of unmeasured state.
\end{abstract}

\section{Introduction}
Specific aspects of dynamics of general entities ---not necessarily 
from the  microphysical domain--- can be successfully described by an 
abstracted formulation of quantum mechanics (Aerts, 1982, 1983ab, 1986, 1992, 1994, 1999). This 
approach  is used here to seize some of the specific dynamical 
aspects of {\em conceptual entities} (Aerts, Broekaert, \& Gabora, 1999, 
2000, 2002).

The  theoretical model which conceives a {\em 
conceptual entity} as  a consistently testable configuration of 
properties in the sphere where personal and interpersonal cognitive 
interactions are taking place, is drawn from analogy with e.g. 
modeling of social entities in a social sphere, or quantum entities 
in the quantum sphere.
Here  conceptual entities are located in their 
proper `space': the cognitive layer  or sphere of reality.  The 
nature of the conceptual entity and its interactions  limits the 
analogy with physical or social modeling; essentially the transience 
of their identity and subjectivity in interaction, and their 
ontological status is different  (Aerts, Broekaert, \& Gabora, 1999, 
2000).

Specifically, in the  application of the abstracted quantum 
formalism to cognition,  the interaction between the  {\em context} 
and a conceptual entity can be modeled. The context is the set of 
effective extraneous factors from physical surrounding and internal 
cognitive state as well. The latter is considered  as the 
ever-fluctuating associative structure of the conceptual network of 
the mind. At present we envisage our `capacity of  logical inference' 
to figure as a {\em coercive} internal context. Which supposes that 
outcomes of  the process of logical inference are endorsed as valid 
cognitive `input' states for further reasoning.

The key factor of 
quantum mechanics that allows its application in the present modeling 
problem, is the contextual and indeterministic effect of the {\em 
measurement} process.  Aspects of formal equivalence between 
abstracted quantum mechanics and the `concept in context'  cognitive 
model  for the Liar Paradox will allow to cast the latter in a strict 
quantum-like model.

We suppose the cognitive entity of the Liar 
Paradox can be validly accessed using language. The most elementary 
form of the Liar Paradox may well be the natural linguistic 
expression ``this proposition is false".   Classically its reasoning 
by logical inference  leads to  the well known logically 
contradictory evaluation, leaving indefinite its genuine logical 
state.  Repetitive reasoning on the Liar Paradox sets in an 
oscillatory attribution of contradicting truth values.

The 
quantum-like  model of Liar Paradox allows  the non-deterministic 
contextual actualization of  logical truth-values and the continuous 
deterministic evolution by reasoning  at any subsequent instance  of 
{\em time} as well.

We develop the formalism required for the Liar 
Paradox model in the next section.


\section{The Quantum Model of the Liar Paradox}
We let the  Liar 
Paradox  be  a configuration of a number of sentences 
---propositions--- referring  to each other and claiming truth or 
falsehood of its target sentence in the configuration.  The entity is 
stripped to its logical content  using a formal shorthand notation. 
In this notation the simplest Liar Paradox is 
\begin{eqnarray} 1 & 
& \not 1 \label{1LP}
\end{eqnarray}
The first number is the sentence pointer. The second 
expression in row is the semantical content of the proposition,  here 
the target sentence  pointer number with a logical operator acting 
upon it (True $\equiv {\bf 1}$ or False $\equiv \not$ \ ).

Some 
8-Liar Paradoxes become in this notation;
\begin{eqnarray}
\begin{tabular}{ccc|ccc}
1 &        $\not 2 $         & & &     5      &       6   \\
2 &           3          & & &     6      &       7  \\
3 &          4               & & &     7       &      8    \\
4 &          5               & & &     8       &      1
\end{tabular}
&{\rm or }&
\begin{tabular}{ccc|ccc}
1 &        $\not 3 $         & & &     5      &  $ \not 1$  \\
2 &        $\not 7$          & & &     6      &   $ \not 5$  \\
3 &          8                   & & &     7       &  $\not 4$ \\
4 &          6                   & & &     8       &      2
\end{tabular} \ \ \ \ {\rm or }\ \ \ \ \ \  ... \label{8LP}
\end{eqnarray}
In the following we will refer to (\ref{8LP},b) as an 
explicit example case. The formulation of the generalized liar 
Paradox is one single and ordered or unordered string of $m$ 
concatenated sentences (`daisy-chain' configuration);
\begin{eqnarray}
1 &\ \ & O_1 (2)        \nonumber \\
2 &\ \ & O_2 (3)      \nonumber \\
  & ... &       \ \ \ \ \ \ \ \ \ \  \ \  \ \ \ \ \ \ \ \ \ \  \ \  \ 
\ \ \ \ \ \ \ \ \  \ \   {\rm (`ordered')} 
\label{basicnLPconfiguration} \\
m  &\ \ & O_m (1)        \nonumber
\end{eqnarray}
Where $O_i$ is at choice  one of the logical operators {\bf 1} or 
$\not {}\ $ . In the basic configuration the sentence with pointer 
$m$ semantically  leads back to the initial sentence with pointer 
$1$.   Unordered configurations ---considered as the result of a 
basis transformation--- are:
\begin{eqnarray} 
1 \leq i, j \leq m, \ \ \ \ \ \ i &\ \ & O_i (j)  \ 
\ \ \ \ \ \ \ \ \  \ \  {\rm (`unordered')} 
\label{variantnLPconfiguration}
\end{eqnarray}
where the set $\{ (i,j)\}$ is a permutation of the  number basis of 
the set$\{ (i, i+1) \}$. 
We require all configurations to have 
sentence pointers ordered 1 to $m$,   and consider different  those 
which expose reversed reasoning ordering (pointer $\to$ target $\to$ 
pointer $\to$ ...).
 
The  number of such index configurations is 
$(m-1)!$. Each configuration has  m relations to which is attributed 
a logical operator {\bf 1} or $\not {}\ $. Paradoxical configurations 
require an uneven number $k$ of  $\not {}\ $-operators ($1\leq k \leq 
m$). As $k$ indistinguishable items can be allocated to $m$ relations 
in $\frac {m!}{k !(k-m)!}$ manner, the total number of $m$ sentence 
Liar Paradox configurations is $(m-1)!\sum_{k=1}^{\leq m} \frac 
{m!}{k !(k-m)!}$ (k uneven).  The particular choice of configuration 
does not affect in any manner the general structure of the model, it 
merely changes the contingencies of the entity's  dynamics.

In the 
next section we approach the construction of the model in three 
parts, i) the representation space and reasoning acts, ii) step 
evolution in particular configuration, iii) continuous 
evolution.

\subsection{Modeling of the Representation Space and 
Reasoning Acts}

{\em Representation space}

\noindent
All sentences, components of the Liar Paradox, are  equivalently 
described  by sub-space vectors. The state of the
Liar Paradox is represented by tensor products of state vectors of 
the sentence sub-spaces. To each sentence in the configuration two 
possible truth-falsehood values by {\em hypothesis}, and an a priori 
unknown number of
truth-falsehood values  by{\em inference} can be assigned.  And each 
sentence state vector is attributed a sufficient number of dimensions 
such that reasoning dynamics occurs without degenerescence of states. 
That is,  all substates  produced by reasoning the Liar Paradox are 
unique and should  occur only once during the completion of a 
reasoning cycle.  We introduce hereto a Hilbert space with the 
minimum number of dimensions required to symmetrically embed the 
$m$-sentence configuration.  The requirement of symmetrical 
representation  reflects the equivalence of all sentences in the 
configuration.
E.g. for the case $m=2$, the symmetrical representation needs the 
Hilbert space {\num C}$^4\otimes
${\num C}$^4$  ( cf. the `double Liar Paradox'  in  Aerts, Broekaert, \& 
Smets, 2000). 
Let us suppose now that $m$ sentences are in the Liar 
Paradox configuration, and let each sentence be represented by an 
$n$-dimensional subspace. The representation space   is then the 
tensor coupled Hilbert space $\Sigma$.
\begin{center}
$\Sigma  \ = \ ${\num C}$^n  _{(1)}  \, \otimes \, ${\num C}$^n 
_{(2)} \, \otimes \, ... \, \otimes \, ${\num C}$^n
_{(m)}$
\end{center}

\noindent
{\em Initial state}

\noindent
We consider the logically 
indefinite conceptual entity of the Liar Paradox as the initial 
situation of the reasoning process. The model then initially is in 
the {\em unmeasured} state with inexplicit truth value due to 
superposition of  state with logically contradicting truth values. It
is then in a state of time-invariance as each component is equally 
undetermined; the conceptual entity Liar Paradox is `cognitively 
perceived but not logically evaluated'. This leads to the constraint 
of imposing an equiponderate initial state in the model.

For 
determination of the subspace dimension $m$ in the appendix, we 
consider the expression for the initial state $\Psi_0$ (for 
generality  the equiponderate demand has not been explicated):
\begin{eqnarray}
\Psi_0 = \sum_{i_1 = 1}^{i_1 = n} ...\sum_{i_m = 1}^{i_m = n} 
\alpha_{i_1 ...i_m} {\bf e}_{i_1 ... i_m}
\label{psizerogeneral}
\end{eqnarray} 
where we systematically employ a double indices 
convention; the first index points at the entry level in a sentence 
state
function, while the number of a sentence itself is indicated by the 
second subindex. E.g., $\alpha_{2_3} =1$
indicates the state function of sentence 3 has `1' in its 2-nd entry. 
The normalization condition for $\Psi_0$ is:
\begin{eqnarray}
  \sum_{i_1 = 1}^{i_1 = n} ...\sum_{i_m = 1}^{i_m = n} \alpha_{i_1 
...i_m} \alpha_{i_1 ...i_m}^* &=& 1  \nonumber
\end{eqnarray}

\noindent
{\em Representation of reasoning acts}

\noindent
The truth and falsehood `measurements' ---reasoning acts--- on each 
sentence correspond to appropriately chosen projectors. These
projectors put the prior state into a state representing truth or 
falsehood by hypothesis of the respective sentence.
Each truth-falsehood by hypothesis projector  represents a possible 
onset of the reasoning on the paradox, as the reasoning
on it can start at any index. The subsequent reading, with logical 
inference, fix the truth-falsehood by inference state of
the remaining sentences in the product. The next step of the dynamics 
is achieved by endorsing the inferred truth value into a
hypothesized value. The sequential appearance of the {\em 
eigenstates} of the  truth-falsehood by hypothesis measurements on the
given sentences is thus realized. The dynamical quantum evolution 
reconstructs the inference sequence.

Without prior specification 
of the dimension, we define for each sentence with pointer
$i$ two projection operators.
  The truth by hypothesis projection operator on sentence $i$ can in 
general be written as:
\begin{eqnarray} T_i &=& \sum_{j=1}^{j=n} {\tau_j}_i \ {\bf 1}_1 
\otimes ... {\bf 1}_{i-1} \otimes P_{\jindex} \otimes {
\bf 1}_{i+1} ...\otimes {\bf 1}_{m}
\label{Tprojector}
\end{eqnarray} and the falsehood by hypothesis 
projection operator on sentence $i$:
\begin{eqnarray} F_i &=& \sum_{j=1}^{j=n} {\phi_j}_i \ {\bf 1}_1 
\otimes ... {\bf 1}_{i-1} \otimes P_{\jindex} \otimes {
\bf 1}_{i+1} ...\otimes {\bf 1}_{m}
\label{Fprojector}
\end{eqnarray} The projectors operates strictly on the subspace with 
same sentence index. The basic projection operators $P_j$ fulfill the 
usual requirements:
\begin{eqnarray} P_j =  P_j^2  &{\rm and}&  P_j =  P_j^\dagger  \nonumber
\end{eqnarray} which leads to, $\forall i, j$;
\begin{eqnarray} {\tau_j}_i, {\phi_j}_i  \in  \{0,1\}  \label{taueq}
\end{eqnarray}
Specific choices of coefficients (\ref{taueq}) on the 
projectors will allow delineated interpretation per entry in the 
vector.

In the next section $n-2$ `truth-falsehood by inference' 
projectors ---similar to (\ref{Tprojector}, \ref{Fprojector})--- for 
each sentence are introduced in order to fulfill the complementarity 
of `false' and `true' operators according Kirchoff and von Neumann 
axioms.  

We settle now the issue of the dimension of all subvectors 
in the $m$ sentence configuration.  When reasoning  the Liar Paradox 
over one cycle, the possible degenerescence of occurring states is 
avoided by supplying sufficient  dimensions $n$ to each subspace. 
The initial state (\ref{psizero}) is spanned over all $n^m$ states, 
while strictly there are  $2 m$ relevant states for the reasoning 
process. For, outcomes of acts of logical inference are endorsed as 
valid `by hypothesis'  input for further cognitive  acts.  I.e.,  $m$ 
outcomes for truth by hypothesis projections and $m$  relevant 
outcomes for falsehood by hypothesis projections on $\Psi_0$.  The 
projection outcomes of the $2m$ truth-falsehood projectors on 
$\Psi_0$  set $2m$ constraints on the model. In the Appendix we prove 
that in order to satisfy $2m$ well chosen constraints on the system 
of $m$ sentences, the subspace for each sentence  needs $n$ 
dimensions, with:
\begin{eqnarray} 
n = 2 m \label{vgldim}
\end{eqnarray}
The model of the Liar Paradox  entity is therefore constructed in a 
$( 2 m )^m $ dimensional Hilbert space.

\subsection{Representation of Evolution: 
Stepwise Reasoning} 
The logically subsequent eigenstates  of the 
reasoning acts ---a truth-falsehood by hypothesis state in product 
relation with truth-falsehood by inference states--- 
following any 
initial measurement, must be  reproduced by dynamical 
evolution.
Therefore at  discrete moment ---indexed $j$ , $1\leq 
j\leq2m$--- of the time-ordering parameter,  $t_j = j \frac{\pi}{2}$, 
at which an inferred logical value is  endorsed into a hypothesized 
logical value, the state vector of the Liar Paradox should be of the 
form, modulo an irrelevant factorizable phase 
$\theta_j$:
\begin{eqnarray}
\Psi \left(t_j\right) = e^{i \theta_j} {\bf e}_{k_1(j)... k_i(j) ... 
k_m(j)}  \nonumber
\end{eqnarray}  
where $\{k_1(j)... k_i(j) ... k_m(j)\}$ are  the 
indices of the tensor product state at step $n$, out of $2m$, of 
reasoning on a
specific Liar Paradox sentences configuration $\{O_i (j)\}$ with a 
truth-falsehood projectors convention (e.g. \ref{taueq}).

Logical reasoning acts  put no conditions on the state functions at 
intermediary  values of the time ordering
parameter, but the quantum formalism allows the integration of the 
stepwise reasoning acts into a continuous evolution (next 
subsection).

The stepwise reasoning dynamics is constructed in the isomorphic 
single Hilbert space, instead of the  tensor coupled Hilbert space 
representation. The transition is done by providing unequivocal 
translation of states.  The basis vectors of the single Hilbert space 
with dimension $n^m$  (eq. \ref{vgldim}) have the index function:
\begin{eqnarray}
\kappa(i_1, ... , i_m) & = &  (2m)^{m-1} (i_m -1) + 
(2m)^{m-2}(i_{m-1} -1)  \nonumber \\ & &   + ... + 2m ( i_2 -1 ) +
(i_1 - 1) + 1  \label{vglindexfunction}
\end{eqnarray}
as a function of  the indices of the basis vectors in the tensor 
coupled representation. The inverse function
$\kappa$ is related to the decimal expression of the $2m$-based digit 
sequence $i_1 i_2 ... i_m$:
\begin{eqnarray} i_1 ...i_m  & = & \left( \kappa (i_1, ...,i_m ) -1 
\right)\vert_{2m-{\rm base}} + 1...1
\label{kappainverse}
\end{eqnarray} 
Where the expression  $1...1$ has $m$ digits. We 
choose the truth and falsehood by hypothesis operators as :
\begin{eqnarray}  T_i &=&  {\bf 1}_1 \otimes ... {\bf 1}_{i-1} 
\otimes T \otimes { \bf 1}_{i+1} ...\otimes {\bf 1}_{m}
\nonumber  \\ F_i &=&  {\bf 1}_1 \otimes ... {\bf 1}_{i-1} \otimes  F 
\otimes { \bf 1}_{i+1} ...\otimes {\bf 1}_{m}
\label{vgloperators}
\end{eqnarray} with
\begin{eqnarray} 
T = 		 
{\tiny \left( \begin{array}{cccc} 
0   &  ... & 0   & 0    \\ 
.. &  ... & ... & ...  \\ 
0   &  ... & 
1   & 0    \\       
0   &  ... & 0   & 0    \end{array} \right)}_{2m 
\times 2m}			 
&{\rm and} &  
F =
{\tiny \left( \begin{array}{cccc}  
0   &  ... & 0   & 0    \\
  ... &  ... & ... & ...  \\
0   &  ... & 0   & 0    \\
0   &  ... & 0   & 1    \end{array} \right)}_{2m \times 2m}   \label{TFop}
\end{eqnarray} 
Then, in each component vector of a sentence, the 
`truth  by hypothesis' state property to corresponds  entry with 
index $2m-1$, and `falsehood by hypothesis' state property 
corresponds to entry with index $2m$. The `truth and falsehood  by 
inference'  operators  are defined in direct relation to the 
assignment of the $2m-2$ remaining entries.

For a given  $m$ sentence configuration having  a {\em reasoning sequence} --- for 
example in (\ref{8LP}) $\{1_T, 3_F, \\ 8_F, 2_F, 7_T, 4_F, 6_F, 5_T\}$ 
--- i.e. a time-sequence of $2m$ eigenstate product vectors of the 
reasoning by inference states, the   assignment procedure is:
\begin{eqnarray}
\begin{minipage}{11cm} i) open $2m$ vectors of $m$ tensorially 
coupled $2m$ dimensional sentence-vectors, ii) assign in a 
sentence-vector
the value 1  respectively to entry $2m$  when `false', and entry 
$2m-1$ when `true',   iii) start by assuming
sentence 1 is `true', iii) by consecutive inference assign in each of 
the $2m$ tensor coupled states the proper truth or
falsehood entries of the implied sentence-vectors,  iv) for filling 
in the unknown entries unequivocally an {\em ad hoc} rule is supplied;
  assign the value 1, in a consecutive inference order and starting 
from a truth state (position $2m-1$), in the next tensor coupled 
vector
in the same sentence-vector to the position with index equal to 
previous index minus one, v) jump one tensor coupled state if
it has the `false' entry 1 at position $2m$.
\end{minipage}   \label{adhocrule}
\end{eqnarray}

\noindent The choice of truth-falsehood operators and 
the assignment procedure completely and  unequivocally define  the 
state vector. For example this gives for a 8-sentence Liar Paradox 
(\ref{8LP}), in the tensor coupled space representation 
$\otimes_{i=1}^{i=8}$  {\num C}${^{16}}_{(i)}$  the initial 
superposition state $\Psi_0$:
\begin{eqnarray}
&& \Psi_0 = \frac{1}{\sqrt{16}} \left\{  {\bf 
e}_{15.10.8.12.7.13.4.9} \right.   \nonumber\\ 
&&+ {\bf 
e}_{14.9.16.11.6.12.3.8} + {\bf e}_{13.8.7.10.5.11.2.16} + {\bf 
e}_{12.16.6.9.4.10.1.7}  \nonumber \\ 
&&+ {\bf 
e}_{11.7.5.8.3.9.15.6} + {\bf e}_{10.6.4.16.2.8.14.5} + {\bf 
e}_{9.5.3.7.1.16.13.4}    \nonumber \\ 
&&+ {\bf 
e}_{8.4.2.6.15.7.12.3} + {\bf e}_{16.3.1.5.14.6.11.2} + {\bf 
e}_{7.2.15.4.13.5.10.1}   \nonumber \\ 
&&+ {\bf 
e}_{6.1.14.3.12.4.9.15} + {\bf e}_{5.15.13.2.11.3.8.14}+ {\bf 
e}_{4.14.12.1.10.2.16.13} \nonumber \\ 
&& \left.+ {\bf 
e}_{3.13.11.15.9.1.7.12} + {\bf e}_{2.12.10.14.8.15.6.11} + {\bf 
e}_{1.11.9.13.16.14.5.10} \right\}
\end{eqnarray}
with e.g.:
\begin{eqnarray} 
{\bf e}_{15.10.8.12.7.13.4.9} &=& {\bf e}_{15}\otimes{\bf 
e}_{10}\otimes{\bf e}_{8}\otimes{\bf e}_{12}\otimes{\bf 
e}_{7}\otimes{\bf e}_{13}\otimes{\bf e}_{4}\otimes{\bf e}_{9} 
\nonumber \\
{\bf e}_{15}&= & (0 , 0, 0, 0, 0,  0, 0, 0, 0, 0, 0,  0, 
0, 0, 1, 0)^{\rm t} , \ \ \  \dots \end{eqnarray}
 In the reduced 
single Hilbert space of $16^8$ dimensions the initial state  using the
indexfunction (eq. \ref{vglindexfunction}) is given by;
\begin{eqnarray}
\Psi_0 &=& \frac{1}{\sqrt{16}}\left\{ 
{\bf e}_{3917179961} 
+ {\bf 
e}_{3640285992}
+ {\bf e}_{3345566240}
+ {\bf e}_{3210230023} 
\right.\nonumber  \\
& &
+ {\bf e}_{2789681382}
+ {\bf 
e}_{2503940053}
+ {\bf e}_{2217086916}
+ {\bf e}_{1930815155} 
\nonumber\\
& &
+ {\bf e}_{4060403106}
+ {\bf e}_{1642316945}
+ {\bf 
e}_{1355985807}
+ {\bf e}_{1321312894} \nonumber \\
& &
\left.+ {\bf 
e}_{1034981885}
+ {\bf e}_{749633644}
+ {\bf e}_{463306331}
+ {\bf 
e}_{177012042} \right\} \label{psizero}
\end{eqnarray}
In the next subsection we will see the dynamical evolution spans  a 
subspace of only $2m$ dimensions
(the basis vectors of the initial state), the occupation of the space 
by the model is therefore rather
scarce.

We conclude by recapitulating the consistent interpretation  of each 
vector entry in the state functions in the general
$m$-sentence case.

With respect to the ad hoc procedure 
(\ref{adhocrule}), in column $i$, the state function of sentence $i$
distinguishes $2m$ states of outcome typified by the index $j$ 
according: \\

\begin{minipage}{11 cm}
\begin{description}
\item[$ 1\leq j \leq m-1$],  ``Sentence  $i$ is true by inference 
according to its
referent sentence" and ``Sentence $i$ is made hypothetically true 
after $j$ inferences"
\item[$ m\leq j \leq 2(m-1)$], ``Sentence  $i$ is false by inference 
according to its
referent sentence" and ``Sentence $i$ is made hypothetically false 
after $j+1 - m $ inferences"
\item[$   j = 2m-1$], ``Sentence  $i$ is true by hypothesis"
\item[$   j = 2m$], ``Sentence  $i$ is false by  hypothesis"
\end{description}
\end{minipage}\\

\bigskip
\noindent Where the referent sentence of $i$ is the sentence implying 
$i$, e.g. in Liar Paradox (\ref{8LP}, b), `1' is the
referent sentence of `3'.

The respective projectors related to the detailed outcome states are 
simply the $2m\times 2m$ diagonal matrices with all
elements zero, except unity at position $(j,j)$.


How does one `measure' on the quantum model  of a Liar Paradox? The 
reasoning on a Liar Paradox consists of two
part-processes, `reading the sentence and inferring a sentence's 
truth or falsehood' according the intensional
semantics of the subject sentence and `hypothesizing', with eventual 
prior knowledge, truth or falsehood on that sentence. The
reasoning process `compulsory' continues by the repetition of this 
reading-inferring and hypothesizing act on the consecutive
sentences. The initial reasoning starts by hypothesizing the truth 
value of a given sentence.

This means that in our description the Liar Paradox within the 
cognitive layer of reality is ---before the
measurement--- not in a predictable true or false state. The `true 
state' and the `false state' of the sentence are
specific states; eigenstates of the measurement projectors 
(\ref{TFop}). In general, the state of the Liar Paradox is not
one of these two eigenstates of a sentence.   Due to the act of 
measurement, and in analogy with what happens during a
quantum measurement, the state of the sentence changes (`collapses') 
into one of the two possible eigenstates, the `true
by hypothesis state' or the `false by hypothesis state'. This act of 
making a sentence true or false can be specifically
described as `read the sentence, make the logical inference and 
hypothesize its truth or falsehood'. The compulsory
consecutive reasoning is represented by the discrete unitary 
evolution operator which evolves a given state of sentence into
its logically reasoned consecutive state.

We will expose this scheme for the 8-Liar Paradox (\ref{8LP},b), and 
see that an initial measurement
followed by the sequence of logical inferences puts into work an 
oscillation dynamics that we can describe by a stepping
evolution matrix, and eventually  by a Schr{\"o}dinger evolution over 
reasoning-time.

All discrete steps of reasoning on the generalised Liar Paradox of 
type (\ref{basicnLPconfiguration}) and
(\ref{variantnLPconfiguration}) can be represented by a discrete $2m 
\times 2m$ evolution matrix
$U_D$.
The matrix $U_D$  is conceived as a step matrix;  i.e. with exactly one 1
on each row and column and all other elements identically zero. Such 
a step matrix is always equivalent to a  basis
transformation  of the matrix with the elements of the lower 
off-diagonal and element $\{1, 2m\}$ equal to 1, or a similar one 
built on the higher off-diagonal matrix.
\begin{eqnarray} 
{U_D}\vert_{\rm sub} &=&  \left.{\tiny \left( 
\begin{array}{ccccc}  
0 & 0 & ... & 0 & 1 \\ 
1 & 0 & ... & 0 & 0 
\\
0 & 1 & ... & 0 & 0 \\
.. &  ...   & ... &  ... &  ...  \\
0 & 0 
& ... & 1 & 0
   \end{array} \right)}\right\vert_{\rm base\ permutation} \label{stepmatrix}
\end{eqnarray}
In the  8-Liar Paradox example (\ref{8LP}) the discrete evolution 
submatrix, governing the reasoning evolution of
sentences  in both the tensor coupled  and  embedded description, is given by;
\begin{eqnarray} 
 {U_D}\vert_{\rm sub} &=&
 {\tiny\left( 
\begin{array}{ccccc} 
0         & \mathbf{1}_{6 \times 6}       & 
0         &          \mathbf{0}_{6 \times 7}        &      0 
\\ 
\vdots &             \vdots                         & \vdots   & 
\vdots                           &  \vdots          \\ 
0         & 
\dots                                     &    0         & 
\dots                  &      1               \\ 
\vdots & 
\mathbf{0}_{7 \times 6}     & \vdots    &  \mathbf{1}_{7 \times 7} 
&   \vdots         \\ 
1         &  \cdots 
&     0         &                \cdots                         & 
0             \\ 
0         &  \cdots 
&     1         &                \cdots                         & 
0             \\ 
 \end{array} \right)}
\end{eqnarray} Notice the submatrix (index `sub') discards all 
trivial dimensions from the description; only a
$2m$-dimensional subspace is actually employed in the evolution of 
$m$-sentences.  This procedure can be applied in any
finite dimension. While the $m$-sentence system has an exponentially 
increasing dimension of its description space, the
relevant dynamics is still only taking place in a $2m$-dimensional 
subspace, i.e. increasing linearly with the number of
sentences.


\subsection{Representation of Evolution:  Continuous Reasoning} 
The 
reasoning on the $m$-sentence Liar Paradox is characterized by 
discrete moments of accumulated inferences-hypotheses
till the completion of the full entity and observation of a 
contradictory truth-value for the initial sentence. The
discreteness in the temporal process features explicitly in the 
logical reasoning; the evolution is characterized by
the completion of consecutive inferences-hypotheses. The formalism of 
operational quantum mechanics allows a continuous time
parameter of evolution. The introduction of continuous time in the 
model allows interpretation of intermediate states
and qualitative duration of the reasoning on the Liar Paradox. Given 
the simplicity of the formal model no strict
relation with psychological time is intended.

For the formal description at every instance of the time-ordering 
parameter, a procedure of diagonalization on the
$2m \times 2m$ submatrix $U_D$ is performed.
\begin{eqnarray}
  U_{D} |_{\rm diag} = R  U_D   R^{-1}
\end{eqnarray}
where $R$ is the $2m \times 2m$ diagonalization matrix.

The diagonalization procedure allows to solve the equation of the
Schr\"odinger evolution operator with Stone's Theorem ---at ordering 
parameter equal to 1 unit of time--- for the
Hamiltonian:
\begin{eqnarray}
  H_{\rm sub} |_{\rm diag} =i \ln  U_D|_{\rm diag}
\end{eqnarray}   Inverting the procedure of diagonalization, the 
infinitesimal generator of the time-evolution ---the
submatrix hamiltonian--- can be obtained.
\begin{eqnarray}
  H_{\rm sub}=  R^{-1} H_{sub} |_{\rm diag} R   \label{Hsub}
\end{eqnarray} The unitary evolution submatrix can be constructed for 
any time $\tau$ (units of a single step in the
reasoning). The continuity of the one-parameter group of unitary 
evolution operators  allows intermediate moments
of the time ordering parameter:
\begin{eqnarray}
\forall \tau: \ \ \ \  U_{\rm sub} (\tau) &=& R^{-1} e^{i E_k \tau} 
g_{kk}  R  \label{Usubtau}
\end{eqnarray}
where the $E_k$ are the eigenvalues of the hamiltonian and $g_{kk}$ 
is the  $2m \times 2m$ diagonal unit
matrix. The unitary evolution operator (eq. \ref{Usubtau}), the 
truth-falsehood projection operators
(eq. \ref{TFop}), and the initial state $\Psi_0$ (eq. \ref{psizero}) 
allow the continuous time description of
any $m$-sentence Liar Paradox of the type 
(\ref{basicnLPconfiguration}) and (\ref{variantnLPconfiguration}). The
representation of the model in the tensor coupled space is 
straightforward, using the inverse function of the $\kappa$
index (eq.
\ref{kappainverse}).

In order to inspect qualitatively the 
reasoning evolution of the Liar Paradox, we make use of the
truth and falsehood probabilities
$P_{i, T} (\tau)$ and  $P_{i, F} (\tau)$ respectively for each sentence
$i$ given the initial state ---in short hand notation--- of e.g. $1_T$:
\begin{eqnarray} P_{i, T} (\tau) &=& \vert \left< i_T \vert U (\tau) 
1_T \right> \vert^2 \nonumber \\ P_{i, F} (\tau)
&=& \vert \left< i_F \vert U (\tau) 1_T \right> \vert^2
\end{eqnarray} In graphical representation it is easily seen how the 
probabilities evolve over time from a given truth
value to their paradoxical opposite value.

\begin{figure}[h]
\begin{center}
\includegraphics{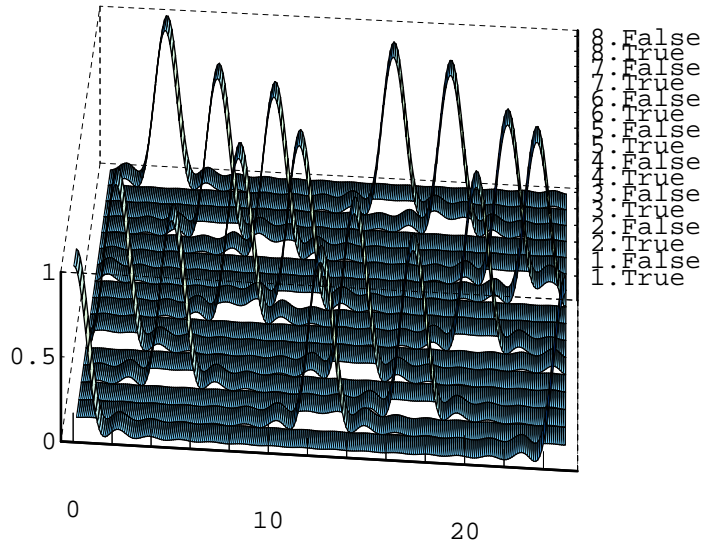}
\caption{{\bf Time evolution} of outcome probabilities for reasoning the 8-Liar Paradox (form. \ref{8LP}, b), with at $t = 
0$, a `true'-measurement
of sentence 1  on the initial state $\Psi_0$. The `time'  $t$ is an
arbitrary continuous ordering parameter without physical
interpretation. Logical contradiction  is apparent after each
interval $\Delta t = 8 \frac{\pi}{2}$. The probability  for a given
outcome ``sentence $i$ is $T/F$  at time $t$" is obtained  by taking
the modulus squared of the sca\-lar product  of sentence of substate
``$i$ is $T/F$" and the  initial state evolved till time $t$, i.e:
$P(i.{\scriptsize T/F}, 1.T, t )=  \left\vert \left< \psi_{i.{\scriptsize T/F}}
\vert U (t) P_{1.T} \Psi_0 \right> \right\vert^2$.}
\label{default}
\end{center}
\end{figure}

\section{Conclusion}
We have treated the Liar Paradox as a special case of a conceptual 
entity: a consistently testable configuration of truth properties 
expressed by sentences, and subject to our capacity of  logical 
inference.  The `contextuality'  of the reasoning process on the Liar 
Paradox  --- here provided by the logical conceptual network in the 
mind --- allows the  construction of an abstracted quantum model. In 
the quantum model an initial  hypothesis on a sentence  engenders a 
time evolution of build up and collapse of logical states and 
eventually logically paradoxical content, without end.   Evidently 
any real world reasoning on  the Liar paradox does not expose this 
compulsory machine-like continuation of the process.  Only the 
capacity of logical inference  --- here as a coercive internal 
context ---  has been accounted for.  Ending the reasoning on the 
sentences needs the hypothesis of reestablishing the original 
superposition state of indefinite logical truth value.

Technically, we have found a procedure to solve in general 
the quantum mechanical modelling problem for the  $m$ sentence Liar
Paradox. The formal model of truth behavior of the Liar Paradox 
needs to be constructed in a  $ (
2 m )^m $ dimensional Hilbert space.  The exponential growth of the 
representation space is due to the demand of symmetric
treatment of component sentences.  The dynamical evolution of the 
Liar Paradox however spans only a subspace of
$2m$ dimensions. The linear dimensional growth of the  relevant space 
allows an adequate description of the
Hamiltonian and unitary evolution operator.  An indefinite state of 
the unreasoned Liar Paradox entity is obtained (eq.
\ref{psizero}) and is time invariant when not reasoned on. The 
time-invariance of the initial state $\Psi_0$ (eq.
\ref{psizero}) --- $\Psi(0) = \Psi(\tau)$ --- follows immediately 
from the fact that it is an eigenvector of the step matrix $U_D$
(eq. \ref{stepmatrix}).

What does the full quantum 
description of the Liar Paradox imply? The crucial feature making the 
Liar Paradox fit to be modeled by abstract quantum mechanics is its 
deterministic swaying to a coercive logical context, provided by the 
conceptual network of the mind.\\
The obtained model extends the 
static `entity + context' configuration by providing a 
phenomenological dynamic. We have therefore been able in
this case to introduce a time-propagator characterizing the 
concatenation of states of thought, albeit restricted to a 
non-evolving  coercive logical context of logical inference.  More 
general conceptual entities with variable internal context and 
environmental context will  most certainly not fit a complete quantum 
model.\\
The appropriated autonomy  of dynamics of the entity does 
not necessarily intend its ontological reality. This reading would be 
a literal interpretation of the physical analogy with the obtained 
complete quantum description. The cognitive person's motivation of 
the entity by reasoning has been the reference for the dynamics of 
the conceptual entity.   The temporal evolution of the entity is 
expected to originate intrinsically, when considered from the quantum 
mechanical analogy, while the  construction mode of the evolution 
supposes the cognitive person's motivation by reasoning.  The 
indistinguishability of the `autonomous evolution of an ontological 
state' and  `coercive evolution by inner logical context'  in the 
model of the Liar Paradox can be interpreted as the mechanism that 
provides its {\em intentionality}. The conceptual entity Liar Paradox 
refers over time to the inner context of logical inference, which
is indeed, in this particular case, sufficient ground for recognizing 
its intentionality.

\begin{figure}[h]
\begin{center}
\includegraphics{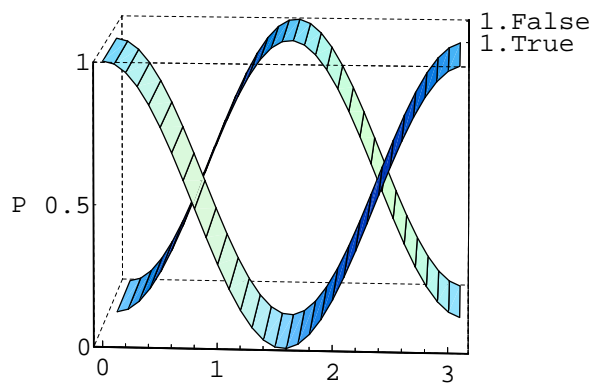}
\caption{{\bf Time evolution} of the 1-Liar 
Paradox (form \ref{1LP}). At  $t = 0$ the entity is prepared in its 
True-state, the False-state is reached at  $\Delta t = 
\frac{\pi}{2}$. Temporal separation of incompatible logical truth 
values circumvents logical paradox.}
\label{default}
\end{center}
\end{figure}

Does the quantum model `solve'  some problems of self-reference in 
the Liar Paradox?  When we attempt to understand the paradox of 
self-reference using classical logical categories, we are caught in a 
logical contradiction. In the present model  this problem is avoided 
by the separation of incompatible logical values over time (Fig.  2). 
The  model  provides intrinsic contextuality to the Liar Paradox, 
distinguishing markedly its nature from conceptual entities with 
predetermined classical truth or falsehood values. The  present model 
therefore suggests that the sphere of conceptual entities includes 
elements not subject to classical logical categories.

\section*{References}

\noindent
Aerts, D. (1982). Description of many physical entities 
without the paradoxes encountered in quantum mechanics.
{\it Foundations of Physics}, {\bf 12}, 1131--1170.

\noindent
Aerts, D. (1983a). Classical theories and non classical 
theories as a special case of a more general theory. {\it Journal of 
Mathematical Physics}, {\bf 24}, 2441--2453.

\noindent
Aerts, D. (1983b). The description of one and 
many physical systems.
In C. Gruber (Eds.), {\it Foundations of Quantum Mechanics} (pp. 63--148). Lausanne: A.V.C.P. 

\noindent
Aerts, D. (1986). A possible explanation for the 
probabilities of quantum mechanics. {\it Journal of  Mathematical 
Physics}, {\bf 27}, 202--210.

\noindent
Aerts, D. (1992). The construction of reality and its 
influence on the understanding of quantum structures.
{\it International Journal of Theoretical Physics}, {\bf 31}, 1815--1837. 

\noindent
Aerts, D. (1994). Quantum structures, separated physical 
entities and probability. {\it Foundations of Physics} {\bf 24}, 
1227--1259.

\noindent
Aerts, D. (1999). Foundations of quantum physics: a general 
realistic and operational approach. {\it International Journal of 
Theoretical Physics}, {\bf 38}, 289--358.

\noindent 
Aerts, D. Broekaert, J. , Gabora, L. (1999). 
Nonclassical contextuality in cognition: Borrowing from
  quantum mechanical approaches to indeterminism and observer 
dependence. {\em Dialogues in Psychology}, 10.0.

\noindent 
Aerts, D. Broekaert, J. , Gabora, L. (2000). 
Intrinsic contextuality as the crux of consciousness. In K. Yasue, M. Jibu \& T. Della Senta (Eds.), {\em 
Proceedings of Fundamental Approaches to Consciousness, Tokyo '99}. Amsterdam: John Benjamins Publishing Company. 

\noindent 
Aerts, D. Broekaert, J. , Gabora, L. (2002). A 
case for applying an abstracted quantum formalism to cognition. In M. H. Bickhard \& R. L. Campbell (Eds.),
{\em Mind in Interaction}. Amsterdam: John Benjamins Publishing Company.

\noindent 
Aerts, D. Broekaert, J., Smets, S. (1999). The 
liar-paradox in a quantum mechanical perspective. {\it Foundations 
of  Science}, {\bf 4}, 115--132. Preprint at \textsf{http://arxiv.org/abs/quant-ph/0007047}. 

\noindent 
Aerts, D. Broekaert, J., Smets, S. (2000). A 
quantum structure description of the liar-paradox. {\it 
International Journal of  Theoretical Physics}, {\bf 38}, 3231--3239. Preprint at \textsf{http://arxiv.org/abs/quant-ph/0106131}.

\par\noindent 
Pitowsky, I. (1989). {\it Quantum 
Probability-Quantum Logic}, Lecture Notes in Physics, {\bf 321}, Springer-Verlag, Wien.

\section*{Appendix: minimal dimension of sentence subspace }
We prove 
the sentence vectors need $n=2m$ dimensions in order not to have a 
degenerescence of states. In general we can write the action of the 
projectors  (\ref{Tprojector}, \ref{Fprojector}) on the initial state 
(\ref{psizerogeneral}), $ 1\leq i \leq m$:
\begin{eqnarray} T_i \Psi_0  &=&  \sum_{j = 1}^{j=n}  \sum_{k_1 =
1}^{k_1 = n} ... \sum_{k_{i-1} = 1}^{k_{i-1} = n} \sum_{k_{i+1} = 
1}^{k_{i+1} = n} ... \sum_{k_m = 1}^{k_m = n}
{\tau_j}_i
\alpha_{k_1 ... \jindex ... k_m} {\bf e}_{k_1 ... \jindex ... k_m} \nonumber
\end{eqnarray} 
Similarly for the false-projectors, $ 1 \leq i \leq m$;
\begin{eqnarray} F_i \Psi_0  &=&  \sum_{j = 1}^{j=n}  \sum_{k_1 = 
1}^{k_1 = n} ... \sum_{k_{i-1} = 1}^{k_{i-1} = n}
\sum_{k_{i+1} = 1}^{k_{i+1} = n} ... \sum_{k_m = 1}^{k_m = n} {\phi_j}_i
\alpha_{k_1 ... \jindex ... k_m} {\bf e}_{k_1 ... \jindex ... k_m} \nonumber
\end{eqnarray}
This choice does not  treat the meaning of the entries 
symmetrically over all sentence vectors in contrast to the choice 
(eqs. \ref{TFop}) . For ease of proof the individual truth-falsehood 
projection operators are chosen such that simple indices occur on
the outcome states.  This can be done due to the equivalence with the 
case where specific
base vector permutations give a unique matrix representation of the individual
projection operator. We consider the constraints:
\begin{eqnarray} T_i \Psi_0  &=& t_i \ {\bf e}_{i ... i ... i} 
\label{Tprojectorofproof} \\ F_i \Psi_0  &=& f_i \  {\bf
e}_{m+i ... m+i ... m+i} \label{Fprojectorofproof}
\end{eqnarray} The coefficients $t_i$, $f_i$ are restricted to: $0 < 
t_i \leq 1$ and $0 < f_i \leq 1$.\\
  Taking into account the explicit expressions of the initial 
superposition state and the projection operators, these
constraints lead to, $i
\in {1, ..., m}$;
\begin{eqnarray} t_i \ {\bf e}_{i ... i ... i}   &=&   \sum_{j = 
1}^{j=n}  \sum_{k_1 = 1}^{k_1 = n} ... \sum_{k_{i-1} =
1}^{k_{i-1} = n} \sum_{k_{i+1} = 1}^{k_{i+1} = n} ... \sum_{k_m = 
1}^{k_m = n} {\tau_j}_i
\alpha_{k_1 ... \jindex ... k_m} {\bf e}_{k_1 ... \jindex ... k_m} 
\nonumber \\ f_i \ {\bf e}_{m+i ... m+i ... m+i}
&=&  \sum_{j = 1}^{j=n}  \sum_{k_1 = 1}^{k_1 = n} ... \sum_{k_{i-1} = 
1}^{k_{i-1} = n} \sum_{k_{i+1} = 1}^{k_{i+1} = n}
... \sum_{k_m = 1}^{k_m = n} {\phi_j}_i
\alpha_{k_1 ... \jindex ... k_m} {\bf e}_{k_1 ... \jindex ... k_m} \nonumber
\end{eqnarray} And due to the orthogonality of the base vectors:
\begin{eqnarray}
\forall i,j, \kdex:  i \leq m , \jindex = i , \kdex = i\ \ \ \, \ \ \ 
\ \ t_i   &=&    {\tau_i} \, \alpha_{i ... i ...
i} \nonumber \\
  \forall i, j , \kdex :   i \leq m , {\rm not}\ \{  \jindex = i , 
\kdex = i \}  \ \ \ \ 0  &=&    {\tau_j}_i \,
\alpha_{k_1 ... \jindex ... k_m}  \nonumber \\
  \forall i, j, \kdex:  i \leq m , \jindex = m + i ,  \kdex = m + i\ \ 
\ \, \ \ \ \ \ f_i   &=&    {\phi_{m+i}} \,
\alpha_{m+i ... m+i ... m+i}  \nonumber \\
  \forall i, j , \kdex :    i \leq m,  {\rm not}\ \{ \jindex = m + i , 
\kdex = m + i \} \ \ \ \ 0  &=&    {\phi_j}_i \,
\alpha_{k_1 ... \jindex ... k_m}  \nonumber
\end{eqnarray} Where the simplified notation has been used $\tau_i 
\equiv {\tau_i}_i$ and $\phi_{m+i} \equiv
{\phi_{m+i}}_{m+i}
$. These equations have, taking into account condition (\ref{taueq}), 
the unique solution;
\begin{eqnarray}
\forall i, j, \kdex: i \leq m ,  \jindex = i ,  \kdex = i & & 
{\tau_i}  = 1    \nonumber \\
  & & \alpha_{i ... i ... i}     = t_i  \nonumber \\
\forall i, j, \kdex: i  \leq m ,  \jindex = m + i ,  \kdex = m + i & 
& \phi_{m+i} = 1 \nonumber \\ & & \alpha_{m+i ...
m+i ... m+i}  = f_i   \nonumber \\
\forall i, j , \kdex :   \forall i \leq m, {\rm not}\ \{  \jindex = i 
,  \kdex = i \}  & & {\tau_j}_i   =  0  \nonumber
\\ & &   \alpha_{k_1 ... \jindex ... k_m}  = 0 \nonumber \\
\forall i, j , \kdex :  \forall i \leq m,  {\rm not}\ \{   \jindex = 
m + i ,  \kdex = m + i \} & & {\phi_j}_i  =   0
\nonumber \\  & &  \alpha_{k_1 ... \jindex ... k_m}   = 0\nonumber
\end{eqnarray} Having found a solution in a Hilbert space with $(2 m 
)^m$ dimensions, we check whether a lesser
dimension is adequate to represent the model.  Strictly a space with 
$ (2 m )^m -1$ dimensions
should be checked for the consistency of the $2m$ constraints. 
Because individual sentences should not be distinct, we lower
the dimension to the symmetric case  of $   ( 2 m - 1 )^m $ 
dimensions, i.e. assigning $n = 2 m-1$  dimensions to each
sentence subspace.\\ We choose the first $2m-1$ conditions 
identically as in the previous case, and complete the
set with one more constraint where we are obliged to re-introduce at 
least $m$ component base vectors indices, e.g.:
\begin{eqnarray}
\forall i  \in \{1, ..., m\}: & &  T_i \Psi_0  = t_i \ {\bf e}_{i ... 
i ... i} \nonumber \\
\forall i  \in \{1, ..., m-1\}: & &  F_i \Psi_0  = f_i \  {\bf 
e}_{m+i ... m+i ... m+i} \nonumber \\
  & & F_m \Psi_0   =  f_m \  {\bf e}_{1 ... 12} \nonumber
\end{eqnarray}  Actually the inevitable reintroduction of at least 
two indices of base vectors will lead finally to
contradiction in the constraints.\\ The explicit expressions of the 
initial superposition state and the projection
operators in the constraints lead to:\\
$\forall i, j, \kdex: i \leq m$
\begin{eqnarray}
  \jindex = i ,  \kdex = i   & & t_i   =    {\tau_i} \, \alpha_{i ... 
i ... i}        \label{vglInco1} \\
  {\rm not}\ \{  \jindex = i ,  \kdex = i \}  &  & 0  =    {\tau_j}_i \,
\alpha_{k_1 ... \jindex ... k_m} \label{vglInco2}
\end{eqnarray}
$\forall  i, j , \kdex : i \leq m -1 $
\begin{eqnarray}
    \jindex = m + i , \kdex = m + i   &  & f_i =    {\phi_i}_{m+i} \, 
\alpha_{m+i ...
m+i ... m+i} \label{vglInco3} \\
  {\rm not}\ \{  \jindex = m + i , \kdex = m + i \}  &  & 0  =    {\phi_j}_i \,
\alpha_{k_1 ...  \jindex ... k_m} \label{vglInco4}
\end{eqnarray}
and
\begin{eqnarray}
   & & f_m = {\phi_2}_{m} \, \alpha_{1 ... 12} \label{vglInco5} \\
  \forall j, \kdex:  \{k_1,...,k_{m-1},j_{(m)}\} \neq  \{1, ..., 1,2\} 
& & 0  = {\phi_j}_{m} \, \alpha_{k_1 ... j_{(m)}}
\label{vglInco6}
\end{eqnarray} From equation (\ref{vglInco5})  and (\ref{taueq}) we obtain
\begin{eqnarray}
  \phi_{2_m} = 1 \ \ \ \ {\rm and} \ \ \ \  \alpha_{1 ... 12} = f_m 
\label{outcome1}
\end{eqnarray}
  next, consider equation (\ref{vglInco1}) with $i = 1$:  $ {\tau_1} . 
\alpha_{1 ... 1} = t_1 $.  Which implies with
equation (\ref{taueq}):
\begin{eqnarray}{\tau_1} = 1 \ \ \ \ {\rm and } \ \ \ \ \alpha_{1 ... 
1} = t_1 \label{outcome2}
\end{eqnarray}
Finally we put $i=j=k= 1$ and  $k_m = 2$ in equation (\ref{vglInco2}) 
to give: $ {\tau_1} . \alpha_{1 ... 12} = 0 $.
Which implies:
\begin{eqnarray}
  \tau_1 = 0 \ \ \ \ {\rm or} \ \ \ \  \alpha_{1 ... 12} = 0 \label{outcome3}
\end{eqnarray}
Conditions (\ref{outcome1}, b), (\ref{outcome2}, a) contradict 
(\ref{outcome3}). The $2m$ constraints  are therefore
too restrictive for the proposed dimension  $ ( 2 m - 1 )^m $ of the 
Hilbert space.\\ The model of the conceptual entity
will therefore need to be constructed in a  $ ( 2 m )^m $ dimensional 
Hilbert space.

\end{document}